\documentclass[a4paper,twoside]{article}
\usepackage{hyperref}

\usepackage{epsfig}
\usepackage{subcaption}
\usepackage{calc}
\usepackage{amssymb}
\usepackage{amstext}
\usepackage{amsmath}
\usepackage{amsthm}
\usepackage{multicol}
\usepackage{pslatex}
\usepackage{apalike}
\usepackage{algorithm2e}
\usepackage{booktabs}   
\usepackage{multirow}   
\usepackage{siunitx}    
\usepackage[bottom]{footmisc}
\usepackage{SCITEPRESS}     

\begin{document}

\title{Lightweight Self-Supervised Detection of Fundamental Frequency and Accurate Probability of Voicing in Monophonic Music}

\author{\authorname{Venkat Suprabath Bitra\sup{1}\orcidAuthor{0000-0002-9254-2274} and Homayoon Beigi\sup{2}\orcidAuthor{0000-0003-0127-2385}}
\affiliation{\sup{1}Dept. of Computer Science, Columbia University, New York, US}
\affiliation{\sup{2}Dept. of Mechanical Engineering and Dept. of Electrical Engineering, Columbia University, and Recognition Technologies, Inc., New York, US}
\email{venkat.s.bitra@columbia.edu, homayoon.beigi@columbia.edu}
\affiliation{Accepted for publication at ICPRAM 2026, March 2-4, 2026, Marbella, Spain}
}


\abstract{Reliable fundamental frequency ($F_0$) and voicing estimation is essential for neural synthesis, yet many pitch extractors
depend on large labeled corpora and degrade under realistic recording artifacts. We propose a lightweight, fully
self-supervised framework for joint $F_0$ estimation and voicing inference, designed for rapid single-instrument training
from limited audio. Using transposition-equivariant learning on CQT features, we introduce an EM-style iterative
reweighting scheme that uses Shift Cross-Entropy (SCE) consistency as a reliability signal to suppress uninformative
noisy/unvoiced frames. The resulting weights provide confidence scores that enable pseudo-labeling for a separate
lightweight voicing classifier without manual annotations. Trained on MedleyDB and evaluated on MDB-stem-synth ground
truth, our method achieves competitive cross-corpus performance (RPA $95.84$, RCA $96.24$) and demonstrates
cross-instrument generalization.}

\keywords{self-supervised pitch detection, unsupervised pitch detection, fundamental frequency, pitch estimation, resonance, musical timbre transfer, probability of voicing, music synthesis, music analysis, CQT, constant Q transform, DDSP, shift cross-entropy loss, musical instrument modeling, ResNeXt neural network, music information retrieval, MIR}

\onecolumn \maketitle \normalsize \setcounter{footnote}{0} \vfill

\section{\uppercase{Introduction}}
\label{sec:intro}

Estimating the \emph{fundamental frequency} ($F_0$) and whether a signal is \emph{voiced} are core primitives for speech and music analysis and they underpin a wide range of downstream systems in Music Information Retrieval (MIR), audio editing, and neural synthesis \cite{r:beigi-sr-book-2011,Yost2009}. Classical pitch trackers such as YIN \cite{deCheveigne2002}, RAPT \cite{Talkin1995}, and pYIN \cite{Mauch2014} exploit periodicity and heuristics that work well in controlled conditions, but become brittle under realistic acoustic artifacts and domain shift.

Deep learning approaches (e.g., CREPE \cite{Kim2018CREPE}) have substantially improved accuracy on benchmark datasets, yet two practical barriers limit their usefulness for \emph{rapid, customized} audio pipelines. First, most high-performing models are supervised and therefore rely on large, carefully annotated corpora that are expensive to build and difficult to scale across instruments, recording conditions, and playing techniques. Second, models trained on curated data often generalize poorly when deployed on real recordings with bleeding, reverberation, and background noise \cite{Morrison2024CrossDomainPitch,Saxena2024DomainShiftMelody,Geirhos2020Shortcut}. These barriers are especially problematic for modern synthesis workflows, where feature extractors must be adapted quickly to a target instrument or recording setup.

A motivating example is Differentiable Digital Signal Processing (DDSP) and related neural synthesis systems \cite{Engel2020DDSP,Hayes2023DDSPReview}. These frameworks can produce high-quality, \emph{personalized} instruments, but assume reliable $F_0$ and voicing features. Practitioners often aim to build a custom synthesizer from \emph{minutes} of audio rather than hours, yet many current pitch extractors, supervised or self-supervised, are trained on large corpora and are not designed to converge robustly in low-data, single-instrument settings. This mismatch creates a bottleneck: even if synthesis can be personalized from 10 to 15 minutes of audio, the feature extractor may not be able to.

This paper reframes pitch estimation as a \emph{low-data, single-instrument, fast-training} problem. Our goal is to learn an $F_0$/voicing front-end that can be trained efficiently from a small amount of target audio, while remaining reliable on real recordings. To emphasize realistic acoustics, we focus on MedleyDB \cite{r-m:bittner-2014-medleydb}, which contains genuine multitrack recordings and therefore exhibits bleed, room effects, and non-ideal isolation. This setting is materially different from digitally resynthesized or perfectly separated datasets (e.g., MDB-stem-synth \cite{Salamon2017}), where unvoiced segments may be near-silent and interference is minimal. Robustness in MedleyDB is critical if the extracted features are to serve as dependable signals for synthesis.

To eliminate the need for manual pitch labels and to support rapid adaptation, we adopt self-supervised learning (SSL) with a transposition-equivariant objective inspired by SPICE \cite{Gfeller2020SPICE} and PESTO \cite{Riou2023PESTO}. Using a Constant-Q Transform (CQT) representation \cite{Brown1991}, we train the model to satisfy a simple consistency constraint: if the input audio is pitch-shifted by a known interval, the model's internal pitch representation must shift by the same amount. This equivariance ``pretext task'' forces the network to encode pitch structure directly, without supervision, and provides a natural mechanism for learning from small, instrument-specific collections.

Beyond $F_0$, we target \emph{robust voicing detection} under interference. In real recordings, ``unvoiced'' does not imply silence; it often contains residual energy from bleed and noise. We exploit this reality through the Shift Cross-Entropy (SCE) loss: the equivariant objective is minimized only when the pitch representation is stable and unambiguous, a condition that is strongly correlated with genuinely voiced content. We use this stability signal to construct reliable pseudo-labels and train a dedicated voicing classifier, producing an $F_0$/voicing pipeline that is both label-free and resilient to realistic artifacts.

\section{\uppercase{Contributions}}
\label{sec:contrib}

\begin{itemize}
  \item Training exclusively on MedleyDB's \cite{r-m:bittner-2014-medleydb} acoustically challenging multitrack recordings to ensure robustness against realistic noise, bleed, and artifacts.
  \item Two-stage pseudo-labeling method where unsupervised Shift Cross-Entropy loss generates confidence scores for training a lightweight binary voicing classifier without manual labels.
  \item Rigorous cross-instrument evaluation quantifying timbre-independent pitch generalization by training on single instruments and testing on unseen instruments.
\end{itemize}



\section{\uppercase{Methods}}
\label{sec:methods}

We build on the self-supervised, transposition-equivariant learning paradigm of PESTO to estimate both fundamental frequency ($F_0$) and voicing from isolated monophonic instrument stems. Our method is tailored to \emph{single-instrument, low-data} training by (i) a CQT front-end that aligns augmentation with musical transposition, (ii) an efficient ResNeXt1D encoder with a Toeplitz-structured pitch head, and (iii) a two-stage voicing pipeline that converts self-supervised stability signals into pseudo-labels for a lightweight binary classifier. The following subsection details the data, feature extraction, augmentations, architecture, and objectives; the next subsection introduces our EM-based reweighting strategy.

\subsection{Dataset and Feature Extraction}
\label{subsec:data_features}

\subsubsection{Datasets}
\label{subsubsec:datasets}

We train on instrument-specific stems from MedleyDB \cite{r-m:bittner-2014-medleydb}, which contains real multitrack recordings and therefore exhibits realistic interference such as bleed, room reverberation, and recording noise. For evaluation, we use MDB-stem-synth \cite{Salamon2017}, which provides precise, frame-level ground-truth $F_0$ labels derived from MIDI-controlled synthesis. Both datasets provide pre-separated stems, avoiding the need for source separation during training and testing. A detailed comparison between MedleyDB and MDB-stem-synth is provided in Section~\ref{subsec:medleydb_vs_mdbss}.

\subsubsection{Constant-Q Transform (CQT) Input}
\label{subsubsec:cqt}

Each audio clip is converted to a log-magnitude CQT representation. Let a single time frame be $x \in \mathbb{R}^{B}$ and a training batch be $X \in \mathbb{R}^{N \times B}$, where $B$ is the number of CQT frequency bins and $N$ is the number of sampled frames. We use $f_{\min}=\texttt{A0}\approx 27.5\,\mathrm{Hz}$ and $36$ bins per octave (i.e., $B_{\mathrm{st}}=3$ bins per semitone). The transform uses $B=269$ bins, spanning approximately $7.5$ octaves up to $f_{\max}\approx 4.9\,\mathrm{kHz}$, covering the pitched range through $\texttt{C8}\approx 4.2\,\mathrm{kHz}$.

Although the target stems in our experiments predominantly occupy an $F_0$ range of roughly $200$-$1200\,\mathrm{Hz}$, we retain a wider CQT band to provide headroom for transposition-based self-supervision. With $f_s=16\,\mathrm{kHz}$ (Nyquist frequency $=8\,\mathrm{kHz}$), content above $\sim 4\,\mathrm{kHz}$ carries limited harmonic evidence for this $F_0$ range. It is included primarily to reduce edge effects, i.e., to lower the probability that pitch shifts push the predicted mode to a boundary bin and cause representation collapse.

\subsubsection{Transposition Equivariance via CQT Shifts}
\label{subsubsec:transposition}

Self-supervision is defined on pitch-shifted pairs $(x,x')$ extracted from the same underlying audio. A transposition of $\tau$ semitones corresponds to a bin shift $\delta=\tau B_{\mathrm{st}}$. We approximate the shifted view by a discrete shift along the CQT frequency axis,
\begin{equation*}
x'(b) \approx x(b-\delta),
\label{eq:cqt_shift}
\end{equation*}
where $b$ indexes CQT bins. The learning objective enforces \emph{equivariance}: the predicted pitch distribution for $x'$ should match the prediction for $x$ shifted by $\delta$ bins. In practice, shifts are applied with boundary-safe padding (rather than wrap-around) to avoid introducing non-physical energy at the frequency limits.

\subsubsection{Pitch-Invariant Augmentations}
\label{subsubsec:invariant_aug}

To encourage invariance to loudness and background perturbations, we apply independent augmentations to each view. Let $x$ denote the log-magnitude CQT frame and $m=\exp(x)$ its linear magnitude. Define, $\mathcal{U}(a,b)$ as the uniform distribution on $[a,b]$, and $\mathcal{N}(\mu,\sigma^2)$ as a Gaussian (normal) distribution with mean $\mu$ and variance $\sigma^2$.

\paragraph{Random gain.}
We sample $g \sim \mathcal{U}(0.5,1.5)$ and apply an additive offset in log space:
\begin{equation*}
x \leftarrow x + \log(g).
\label{eq:gain_aug}
\end{equation*}

\paragraph{SNR-targeted noise.}
We sample $\mathrm{SNR}_{\mathrm{dB}} \sim \mathcal{U}(15,50)$ and compute signal power $P_s=\mathbb{E}[m^2]$ over bins. With $\mathrm{SNR}=10^{\mathrm{SNR}_{\mathrm{dB}}/10}$, we set $P_n=P_s/\mathrm{SNR}$ and draw $n \sim \mathcal{N}(0,P_n)$:
\begin{equation*}
m \leftarrow m + n,\qquad
x \leftarrow \log(\max(m,\epsilon)),
\label{eq:snr_aug}
\end{equation*}
where $\epsilon$ prevents the logarithm of non-positive values.

\subsubsection{Model Architecture}
\label{subsubsec:arch}

\begin{figure*}[!htbp]
  \centering
  \includegraphics[width=\textwidth]{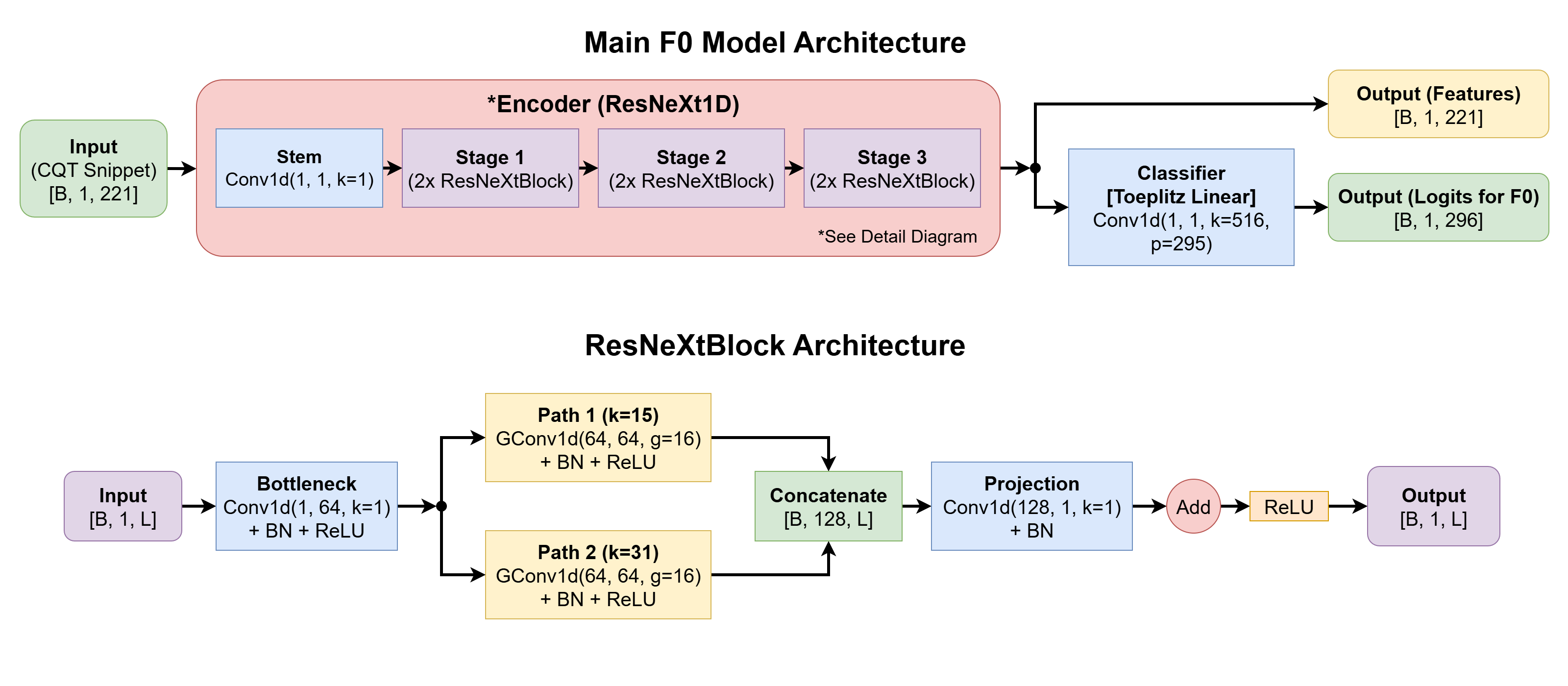}
  \caption{\textbf{Model architecture.} ResNeXt1D encoder over CQT frames followed by a Toeplitz linear classifier (implemented as a 1D convolution) producing pitch logits.}
  \label{fig:architecture}
\end{figure*}

Our network $f_\theta(\cdot)$ maps a CQT frame to pitch logits. It consists of a ResNeXt1D encoder followed by a Toeplitz-structured linear classifier (Fig.~\ref{fig:architecture}). The encoder comprises a stem convolution and three stages of ResNeXtBlock1D modules with grouped convolutions, enabling multi-scale spectral aggregation with improved parameter efficiency. The pitch head is implemented as a 1D convolution over the frequency axis, imposing a Toeplitz structure that reduces parameters relative to a fully connected classifier. This structure also supports efficient post-hoc calibration: a circular shift of the classifier kernel can correct systematic pitch-bin offsets without retraining.

\subsubsection{Training Objectives}
\label{subsubsec:objectives}

Given a batch of paired frames $\{(x_i,x'_i)\}_{i=1}^{N}$ and corresponding logits
$y_i=f_\theta(x_i)$ and $y'_i=f_\theta(x'_i)$, we define
$p_i=\mathrm{softmax}(y_i)$ and $p'_i=\mathrm{softmax}(y'_i)$. We use a weighted sum of three loss terms:
\begin{equation*}
\mathcal{L}_{\text{total}}=
\frac{1}{N}\sum_{i=1}^{N} w_i\Big(
\mathcal{L}_{\text{equiv},i}+
\mathcal{L}_{\text{invar},i}+
\mathcal{L}_{\text{sce},i}
\Big)
\end{equation*}
where $w_i$ are sample weights produced by our iterative reweighting scheme (described in the next subsection).

\subsection{Iterative Re-weighting using Expectation-Maximization}
\label{sssec:reweighting}

\subsubsection{Intuition}
SCE enforces a self-consistency constraint under pitch transposition: if a frame contains a coherent harmonic structure, then the predicted pitch distribution on a transposed view should match a shifted version of the prediction on the original view. Voiced frames approximately satisfy this constraint, yielding aligned targets and low-variance gradients that sharpen the pitch peak. Unvoiced frames (noise, bleed, reverb tails) do not transform into a consistent pitched pattern under transposition, so SCE produces inconsistent targets and high-variance gradients. In low-data regimes, once voiced frames are fit, their SCE becomes small and optimization can become dominated by these uninformative unvoiced gradients. Our EM-style reweighting prevents this by down-weighting frames whose SCE indicates unreliable transposition structure.

\subsubsection{Algorithm (EM-style iterative reweighting)}
\label{sssec:reweighting_algo}
Training alternates between weight estimation (E-step) and weighted optimization (M-step). Let $\mathcal{D}=\{X_i\}_{i=1}^M$ be the training set and initialize $w_i \leftarrow 1$. We update weights every $K=5$ epochs.

\paragraph{Initialization.} Train $f_\theta$ for $K$ epochs with $w_i=1$.

\paragraph{E-step (weight update).} At epoch index $e$ (multiple of $K$), compute $\mathcal{L}_{\mathrm{sce},i}$ for all frames under the current model. Normalize within the pass:
\[
\hat{\mathcal{L}}_{\mathrm{sce},i}=\frac{\mathcal{L}_{\mathrm{sce},i}-\min_j \mathcal{L}_{\mathrm{sce},j}}{\max_j \mathcal{L}_{\mathrm{sce},j}-\min_j \mathcal{L}_{\mathrm{sce},j}+\epsilon}\in[0,1].
\]
Define an epoch-dependent annealing factor
\[
\lambda(e)=\exp\!\left(\frac{e^{1.25}}{1000}\right)-1,
\]
and update weights by
\[
\Delta w_i=\lambda(e)\,\hat{\mathcal{L}}_{\mathrm{sce},i}
\qquad
w_i \leftarrow \max(0,\,w_i-\Delta w_i).
\]
Thus, frames with larger SCE are progressively down-weighted as training advances.

\paragraph{M-step (weighted optimization).} For the next $K$ epochs, update $\theta$ by minimizing the weighted objective. Repeat the E-step and M-step until convergence.

\subsubsection{Why reweighting suppresses bad gradients}
\label{sssec:reweighting_proofsketch}

Let $X \in \mathbb{R}^{B}$ be a CQT frame and $y = f_\theta(X) \in \mathbb{R}^{K}$ be the model's output logits over $K$ pitch bins. Let $X'$ be its transposed view with $y' = f_\theta(X')$. Define $p=\mathrm{softmax}(y)$ and $p'=\mathrm{softmax}(y')$. Let $\text{roll}(y, \delta)$ denote a circular shift by $\delta$ bins (pitch transposition in log-frequency space). The Shift Cross-Entropy (SCE) loss compares predictions on an original frame $X$ and its pitch-shifted version $X'$:
$$
\mathcal{L}_{\text{sce}} = \frac{1}{2} \left( \text{CE}(y', \text{roll}(p, \delta)) + \text{CE}(y, \text{roll}(p', -\delta)) \right)
$$
where $y = f_\theta(X)$ and $y' = f_\theta(X')$. The gradient of cross-entropy with respect to logits follows the standard form:
$$
\frac{\partial \text{CE}(p, q)}{\partial y} = p - q
$$
where $p = \text{softmax}(y)$ and $q$ is the target distribution (treated as fixed).

\paragraph{Clean Silence Encourages Trivial Solutions}
On synthetic datasets (such as MDB-stem-synth), silent frames are exactly zero: $X = \mathbf{0}$. When the input is zero, the model output becomes time-constant: $y = f_\theta(\mathbf{0})$ produces the same logits for all silent frames. With typical initialization, weight decay, and symmetric training data, the network tends to produce approximately uniform predictions on constant inputs, so:
$$
p = \text{softmax}(f_\theta(\mathbf{0})) \approx \mathbf{u}
$$
where $\mathbf{u}$ is the uniform distribution: $$u_k = \frac{1}{K}$$ Since uniform distributions are invariant to circular shifts ($\text{roll}(\mathbf{u}, \delta) = \mathbf{u}$), both the prediction and target become uniform:
$$
\frac{\partial \mathcal{L}_{\text{sce}}}{\partial y} \approx \mathbf{u} - \mathbf{u} = \mathbf{0}
$$

Consequently, the network learns nothing from silent frames and they produce vanishing gradients. This is benign and not informative.

\paragraph{Real Silence is Noisy and Produces Bad Gradients} On real recordings (like MedleyDB), ``silent'' frames are not actually silent. They contain noise, bleed, and reverb: $X = \varepsilon$ (random, nonzero). The deep network $f_\theta$ is highly nonlinear and sensitive to input perturbations. Even small noise $\varepsilon$ can produce significantly different outputs:
$$
y = f_\theta(\varepsilon), \quad p = \text{softmax}(y)
$$
Because the noise varies randomly across frames and over time, $p$ fluctuates unpredictably.

\paragraph{Key Issue} For the pitch-shifted version $X' = \text{roll}(X, \delta)$, there is no consistent pitch relationship between $p(X')$ and $\text{roll}(p(X), \delta)$ because: (i) Noise has no harmonic pitch structure to transpose, (ii) Different noise realizations (e.g., different positions in reverb tails, different microphone bleed patterns) do not maintain any pitch correspondence under transposition, and (iii) The deep network $f_\theta$ may amplify these inconsistencies through its nonlinear layers.

This makes the SCE target $q = \text{softmax}(\text{roll}(y, \delta))$ itself random and inconsistent with the prediction $p' = \text{softmax}(y')$. The gradient variance becomes:
$$
\text{Var}\left[\frac{\partial \mathcal{L}_{\text{sce}}}{\partial y_k}\right] = \text{Var}[p_k - q_k] \approx \text{Var}[p_k] + \text{Var}[q_k]
$$

This leads to high-variance noisy gradients, ``bad gradients'', on unvoiced frames. The model receives contradictory signals across mini-batches, leading to poor convergence, spurious peaks in low-SNR regions and unstable training. 

\paragraph{Solution: EM-Style Reweighting Using SCE}

We treat voicing as a latent variable and use SCE loss as a reliability signal. Frames with low SCE loss have a consistent pitch structure (voiced), while frames with high SCE loss are noisy/unvoiced.
The computed gradients for this weighted loss is given as:
$$
\frac{\partial}{\partial \theta}\big[w_i \cdot \text{CE}(p_i, q_i)\big] = w_i \cdot (p_i - q_i) \cdot \frac{\partial y_i}{\partial \theta}
$$
When $\mathcal{L}_{\text{sce},i}$ is large (noisy frame), $w_i \approx 0$, so the gradient contribution vanishes. This:
\begin{enumerate}
    \item \textbf{Preserves strong learning} on genuinely voiced frames (stable structure $\rightarrow$ low SCE $\rightarrow$ $w \approx 1$)
    \item \textbf{Suppresses high-variance gradients} from noisy/unvoiced frames (unstable structure $\rightarrow$ high SCE $\rightarrow$ $w \approx 0$)
\end{enumerate}
The EM approach turns a liability (noisy real-world silence) into an advantage: the model automatically learns to focus on frames with genuine pitch content while ignoring those dominated by noise.

\subsection{Supervised Voicing Classification}

After self-supervised training, the network $f(\cdot)$ is frozen and the final sample weights $w_i \in [0,1]$ are thresholded at $\theta$ to produce binary pseudo-labels $v_i^* \in \{0,1\}$. A lightweight linear classifier is then trained on CQT frames using binary cross-entropy against these pseudo-labels, producing robust voicing predictions without any manual annotation. Architectural details are provided in Appendix.

\section{Results and Discussion}

\subsection{Recordings vs. Resynthesized}
\label{subsec:medleydb_vs_mdbss}

\begin{figure}[!htbp]
    \centering
    \includegraphics[width=\linewidth]{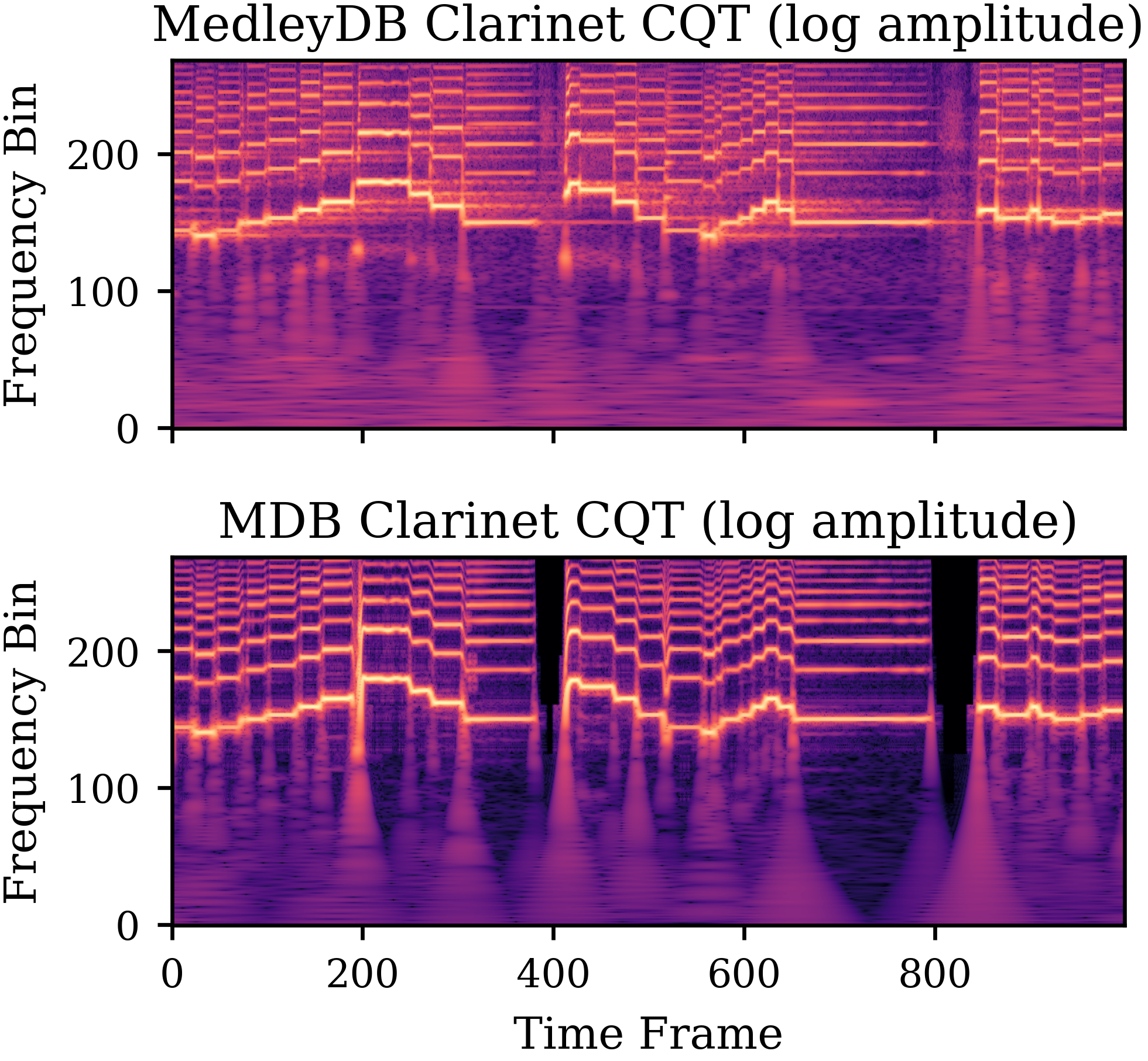}
    \caption{\textbf{Acoustic comparison: MedleyDB vs. MDB-stem-synth.} (Top) MedleyDB recordings show broader partials, modulation sidebands, and exponential decay tails from room acoustics. (Bottom) MDB-stem-synth exhibits narrow harmonic tracks with abrupt terminations and exact silence (black regions). Real recordings retain residual energy from bleed and reverb where synthetic stems are perfectly zero.}
    \label{fig:medley_vs_mdb}
\end{figure}

Figure~\ref{fig:medley_vs_mdb} contrasts CQT magnitude spectrograms from MedleyDB (real multitrack recordings) and
MDB-stem-synth (algorithmic resynthesis) over comparable musical segments. The two corpora differ in ways that directly
affect both $F_0$ tracking and voicing.

MDB-stem-synth exhibits clean harmonic tracks with narrow partial bandwidth and abrupt note offsets, often producing
perfectly silent gaps (exact zeros) between notes. In contrast, MedleyDB contains physically plausible decay tails shaped by
instrument dynamics, microphone placement, and room acoustics. Real performances also include frequency modulation
(vibrato) and amplitude modulation (tremolo), which broaden harmonics and introduce modulation sidebands, visible as
wider ridges around the harmonic ladder.

Finally, MedleyDB includes residual room reflections and inter-track bleed that lift the spectral floor and introduce weak
cross-harmonic artifacts. These effects are absent in MDB-stem-synth, where ``unvoiced'' regions are truly anechoic.
This discrepancy helps explain why models optimized on resynthesized corpora can degrade on real audio: they are never
forced to handle bleed-induced low-SNR frames, realistic decays, or modulation spread. Training exclusively on MedleyDB
therefore targets the operating regime required by real-world synthesis and personalization pipelines.

\subsection{Effect of EM Reweighting on MedleyDB}
\label{subsec:weighting_behavior}

\begin{figure}[htbp]
    \centering
    \includegraphics[width=\linewidth]{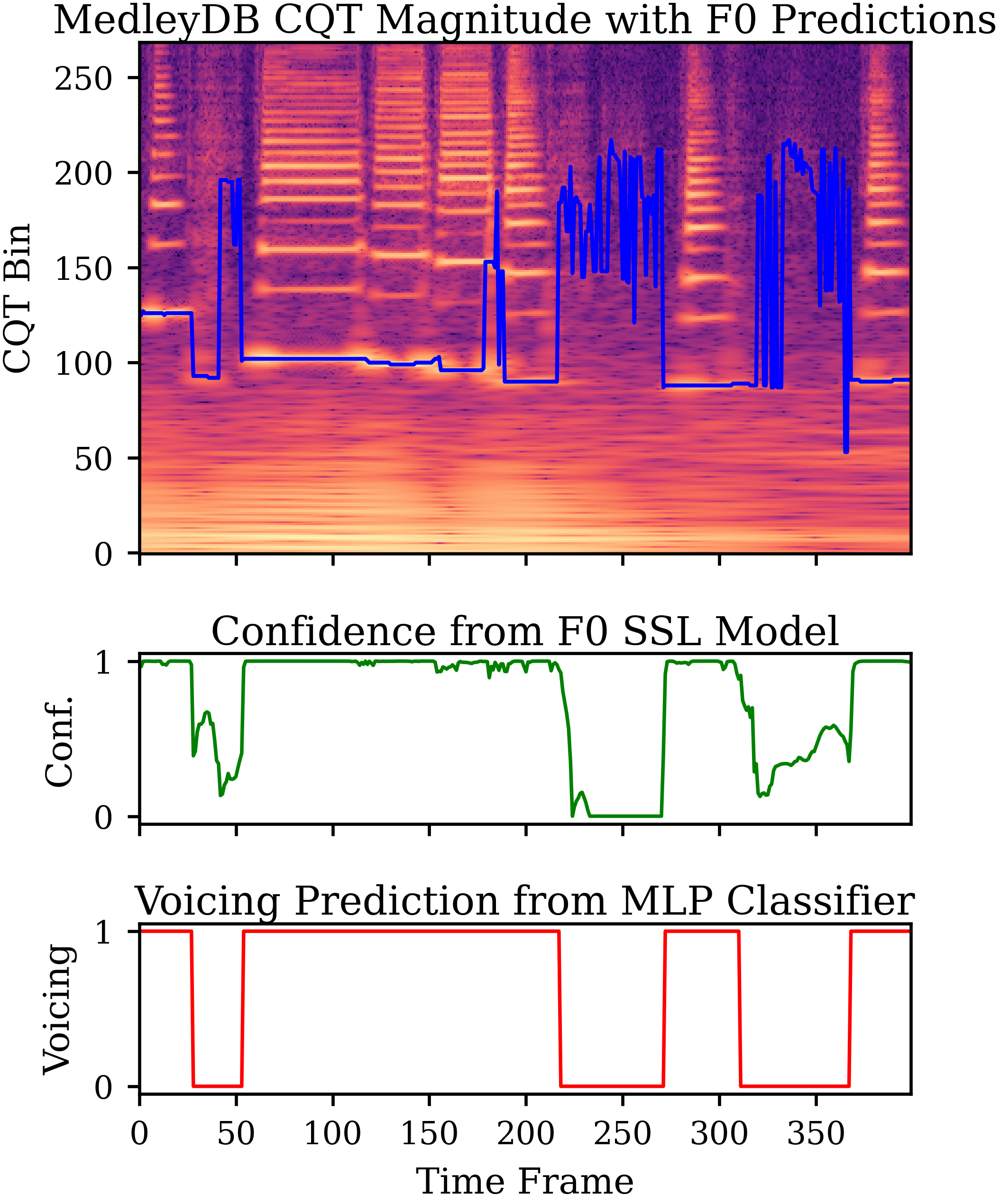}
    \caption{\textbf{Learned weight behavior on MedleyDB.} (Top) CQT spectrogram with predicted $F_0$ (black). (Middle) EM sample weights $w$ (green) remain high on sustained harmonics and drop during releases and transients. (Bottom) Derived voicing $\hat{v}$ (red) effectively gates voiced segments, automatically distinguishing stable pitch from ambiguous content.}
    \label{fig:weights_example}
\end{figure}

Figure~\ref{fig:weights_example} visualizes typical behavior on MedleyDB: the CQT with predicted $\hat{F}_0$ (top), learned
sample weights $w$ (middle), and derived voicing proxy $\hat{v}$ (bottom). We observe a consistent pattern across diverse
recording conditions. During sustained harmonic regions, weights remain high ($w \approx 1$) and stable, and the predicted
$\hat{F}_0$ follows the dominant harmonic ridge. During note releases, reverberant gaps, and ambiguous low-SNR regions,
weights drop sharply, preventing the optimizer from overfitting pitch-unreliable frames.

\begin{figure}[htbp]
    \centering    
    \includegraphics[width=\linewidth]{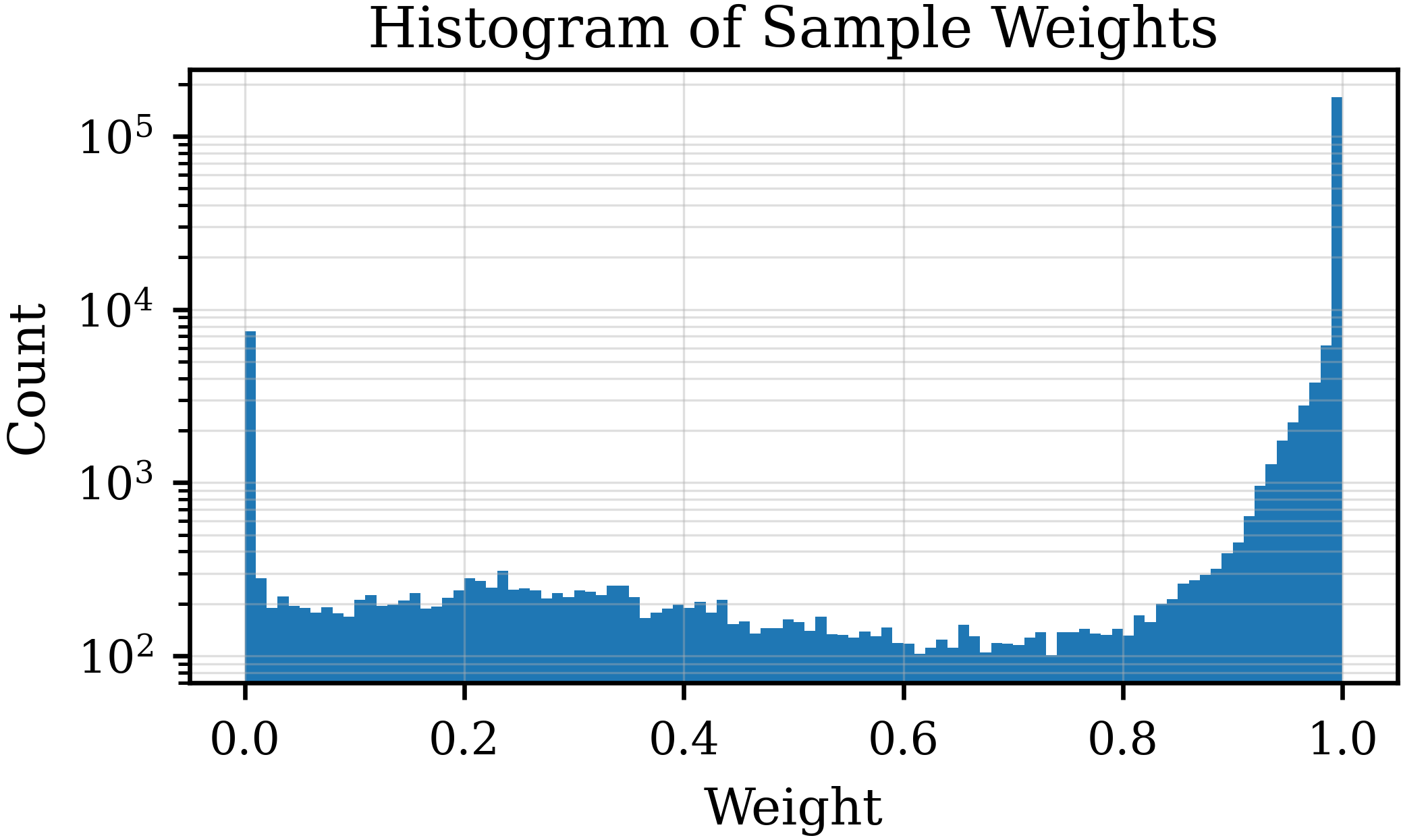}
    \caption{\textbf{Histogram of EM sample weights} on MedleyDB (log-count scale). Clear bimodality: most frames are confidently kept ($w\!\approx\!1$) or down-weighted ($w\!\approx\!0$).}
    \label{fig:weight_hist}
\end{figure}

Figure~\ref{fig:weight_hist} further shows that the learned weights are strongly bimodal (mass concentrated near $0$ and $1$),
indicating that the reweighting naturally separates frames that contain transposition-consistent harmonic structure from
frames dominated by noise/bleed. This bimodality also makes the mapping from $w$ to a voicing estimate $\hat{v}$ stable
without extensive tuning.

\subsection{Cross-Corpus and Cross-Instrument Evaluation}
\label{subsec:cross_corpus}

To quantify transfer under realistic acoustics, we train on MedleyDB stems and evaluate on MDB-stem-synth,  providing
accurate frame-level $F_0$ labels. We report (i) a compact baseline comparison to contextualize our setting, and (ii) a
cross-instrument protocol that trains on a \emph{single} MedleyDB instrument and tests on \emph{unseen} instruments in
MDB-stem-synth. This isolates timbre generalization while holding evaluation labels fixed and precise. It also
reflects the intended use case: rapid, instrument-specific training on real recordings with bleed and reverberation, followed
by deployment on cleaner or differently captured audio.

\paragraph{Baseline context.}
Table~\ref{tab:baselines} summarizes representative supervised (CREPE, DeepF0) and self-supervised (SPICE, DDSP-inv,
PESTO) results reported on MDB-stem-synth. In contrast, our primary setting trains on MedleyDB recordings and evaluates
on MDB-stem-synth labels, introducing a deliberate domain shift. Despite this mismatch, our method remains competitive:
training on MedleyDB and testing on MDB-stem-synth yields $\mathrm{RPA}=95.84$ and $\mathrm{RCA}=96.24$, compared to
$\mathrm{RPA}=96.54$ and $\mathrm{RCA}=97.55$ when trained and tested in-domain on MDB-stem-synth. The small drop
($0.70$ RPA, $1.31$ RCA) indicates that learning under realistic artifacts preserves accuracy on clean evaluation data while
improving robustness in the operating regime. For PESTOv2, the values in parentheses denote our CQT-based interpolation
of the VQT-reported results, included for a fairer comparison under a single time-frequency representation.

\begin{table}[t]
\centering
\renewcommand{\arraystretch}{1.1}
\caption{Baseline comparison on MDB-stem-synth (RPA/RCA in \%). PESTOv2 values in parentheses denotes the results for CQT rather than VQT.}
\label{tab:baselines}
\begin{tabular}{|l|c|c|}
\hline
\textbf{Model} & \textbf{RPA} & \textbf{RCA} \\ \hline
\hline
CREPE \cite{Kim2018CREPE} & 96.7 & 97.0 \\ \hline
DeepF0 \cite{Singh2021} & 98.3 & 98.4 \\ \hline \hline
SPICE \cite{Gfeller2020SPICE} & 89.1 & -- \\ \hline
DDSP-inv \cite{ddspinv} & 88.5 & 89.6 \\ \hline
PESTOv1 \cite{Riou2023PESTO} & 94.6 & 95.0 \\ \hline
PESTOv2 \cite{Riou2025} & 94.4 & 95.0 \\
\hline \hline
Ours (MDB $\rightarrow$ MDB) & 96.54 & 97.55 \\ \hline
Ours (Medley $\rightarrow$ MDB) & 95.84 & 96.24 \\
\hline
\end{tabular}
\end{table}

\paragraph{Cross-instrument transfer.}
Table~\ref{tab:cross_instrument} reports cross-instrument generalization: each column trains on a \emph{single} MedleyDB
instrument, and each row tests on an unseen MDB-stem-synth instrument. Several transfers exceed $90\%$ RPA on the clean
synthetic corpus, consistent with its reduced acoustic complexity (narrower-band partials and exact silences), which makes
harmonic structure more separable. Transfer is strongest within the string family (e.g., violin$\leftrightarrow$cello), reflecting
shared excitation mechanisms and overlapping registers. Bass instruments are the hardest targets: averaged across training
instruments, electric bass attains the lowest mean RPA ($\approx 68.7\%$) and double bass the next lowest ($\approx 77.2\%$),
suggesting that low-frequency dominance and transient-heavy onsets amplify cross-timbre mismatch. Across all conditions,
RCA is consistently higher than RPA, indicating that residual errors are dominated by octave confusions rather than arbitrary
pitch drift, a favorable failure mode for synthesis and conditioning.

\begin{table*}[htbp]
\centering
\caption{Train on MedleyDB (columns), test on MDB-stem-synth (rows). Metrics are \% RPA / RCA.}
\label{tab:cross_instrument}
\resizebox{\textwidth}{!}{%
\renewcommand{\arraystretch}{1.15}
\begin{tabular}{|c|cc|cc|cc|cc|cc|}
\hline
{ \textbf{Instrument Trained $\rightarrow$}} & \multicolumn{2}{c|}{{ \textbf{Clarinet}}} & \multicolumn{2}{c|}{{ \textbf{Cello}}} & \multicolumn{2}{c|}{{ \textbf{Double Bass}}} & \multicolumn{2}{c|}{{ \textbf{Electric Bass}}} & \multicolumn{2}{c|}{{ \textbf{Violin}}} \\ \hline
{ \textbf{Instrument Tested $\downarrow$}} & \multicolumn{1}{c|}{\textbf{RPA}} & \textbf{RCA} & \multicolumn{1}{c|}{\textbf{RPA}} & \textbf{RCA} & \multicolumn{1}{c|}{\textbf{RPA}} & \textbf{RCA} & \multicolumn{1}{c|}{\textbf{RPA}} & \textbf{RCA} & \multicolumn{1}{c|}{\textbf{RPA}} & \textbf{RCA} \\ \hline
{ \textbf{Clarinet}} & \multicolumn{1}{c|}{94.29} & 95.37 & \multicolumn{1}{c|}{94.31} & 95.96 & \multicolumn{1}{c|}{80.61} & 92.68 & \multicolumn{1}{c|}{96.88} & 96.98 & \multicolumn{1}{c|}{94.69} & 95.30 \\ \hline
{ \textbf{Cello}} & \multicolumn{1}{c|}{83.61} & 86.62 & \multicolumn{1}{c|}{83.29} & 91.15 & \multicolumn{1}{c|}{96.34} & 96.65 & \multicolumn{1}{c|}{94.80} & 95.23 & \multicolumn{1}{c|}{87.96} & 90.26 \\ \hline
{ \textbf{Double Bass}} & \multicolumn{1}{c|}{54.33} & 60.45 & \multicolumn{1}{c|}{74.97} & 82.62 & \multicolumn{1}{c|}{92.88} & 93.44 & \multicolumn{1}{c|}{88.48} & 89.68 & \multicolumn{1}{c|}{75.35} & 81.22 \\ \hline
{ \textbf{Electric Bass}} & \multicolumn{1}{c|}{47.22} & 60.44 & \multicolumn{1}{c|}{60.68} & 82.23 & \multicolumn{1}{c|}{89.59} & 91.36 & \multicolumn{1}{c|}{86.22} & 88.56 & \multicolumn{1}{c|}{59.69} & 80.02 \\ \hline
{ \textbf{Violin}} & \multicolumn{1}{c|}{91.50} & 95.37 & \multicolumn{1}{c|}{91.37} & 96.65 & \multicolumn{1}{c|}{86.05} & 91.39 & \multicolumn{1}{c|}{93.48} & 93.85 & \multicolumn{1}{c|}{93.76} & 96.07 \\ \hline
\end{tabular}%
}
\end{table*}

\subsection{Short-Segment Additive Resynthesis}
\label{subsec:additive_resynth}

For qualitative validation, we drive a simple additive synthesizer using the model's framewise $\hat{F}_0$ and voicing
weights $\hat{v}$ (details in Appendix). Figure~\ref{fig:additive_resynth} compares
CQTs of (a) target MedleyDB audio, (b) synthesized output, and (c) MDB-stem-synth audio.

\begin{figure}[htbp]
  \centering
  \includegraphics[width=\linewidth]{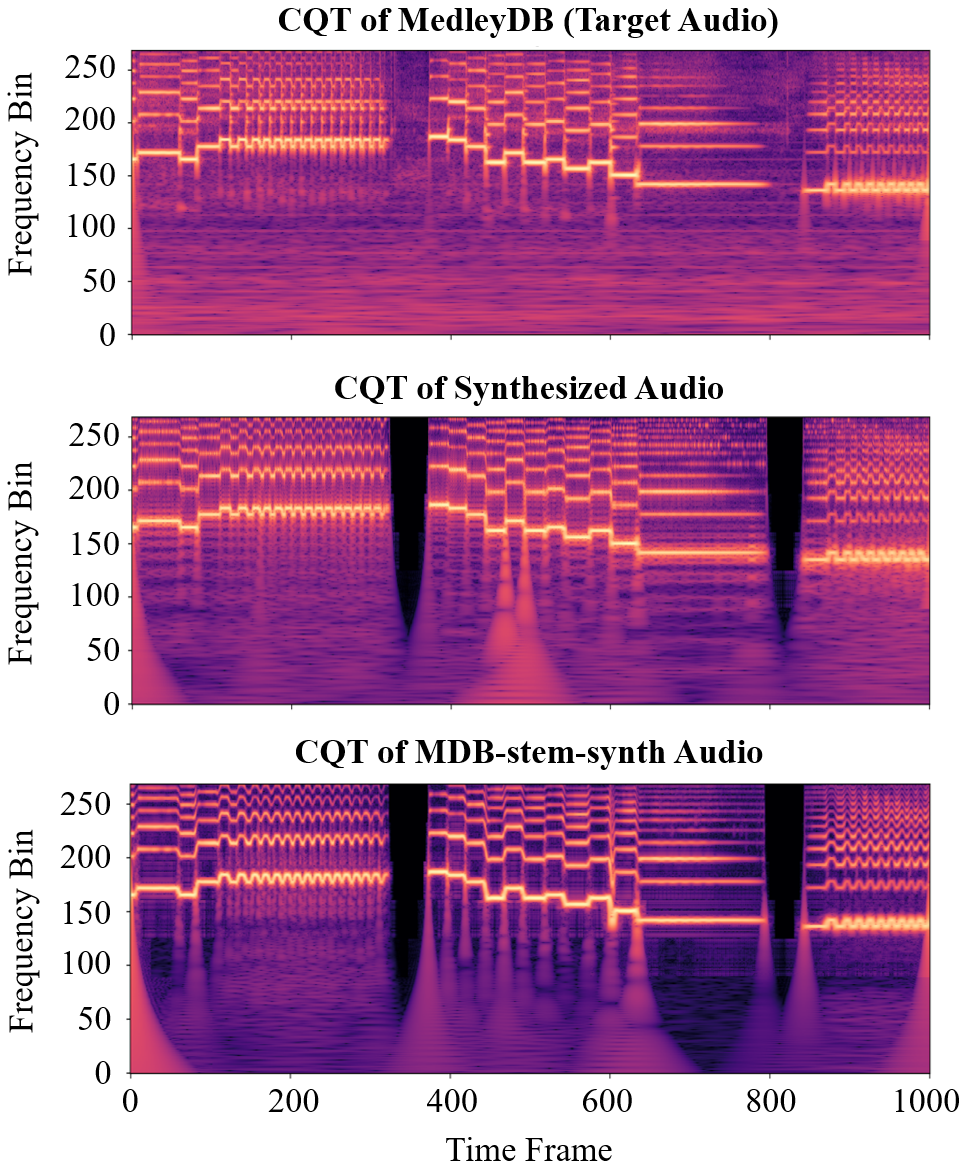}
  \caption{(a) \textbf{Target audio (CQT).} Clear stair-step contours from MedleyDB clarinet with narrow partials and clean inter-note gaps. (b) \textbf{Synthesized audio (CQT) from $\hat{F}_{0}$ and $\hat{v}$.} Harmonic ladders align with the target. The black gaps indicate voicing-controlled muting during predicted unvoiced spans. Low-frequency wedges at the very left/right arise from segment fade-in/out. (c) \textbf{MDB-stem-synth Synthesis (CQT).} Resynthesized audio exhibits narrow, high-contrast harmonic partials with hard on/offsets, exact silences (black gaps), and minimal modulation sidebands compared to real recordings.}
  \label{fig:additive_resynth}
\end{figure}

The synthesized harmonic ladder closely follows the stepwise $\hat{F}_0$ contour of the target, with transitions occurring at
the correct frames, indicating faithful pitch tracking. The voicing gate suppresses ambiguous inter-note regions, producing
clear muted spans in the synthesized CQT and removing residual ringing that persists in real recordings (e.g., release tails).
Minor triangular wedges near segment boundaries arise from overlap-add windowing; they can be reduced with longer fades
but are shown here to make gating locations explicit.

\section{Conclusion}
\label{sec:conc}
We presented a lightweight, fully self-supervised framework for joint $F_0$ estimation and voicing inference from
monophonic instrument stems under realistic acoustics. Training exclusively on MedleyDB exposes the model to bleed,
reverberation, and low-SNR inter-note regions that commonly break pitch trackers trained on resynthesized data.

Our key contribution is an EM-style iterative reweighting strategy that uses SCE-based transposition consistency as a
reliability proxy. Stable harmonic frames retain high weights and drive learning, while unreliable frames are progressively
down-weighted, suppressing high-variance gradients. The converged weights serve as confidence scores for pseudo-labeling
and training a lightweight binary voicing classifier without manual annotations.

Quantitatively, the method remains competitive under cross-corpus transfer
(Medley$\rightarrow$MDB: RPA $95.84$, RCA $96.24$) with only a small gap to in-domain training
(MDB$\rightarrow$MDB: RPA $96.54$, RCA $97.55$), and the cross-instrument matrix shows strong within-family transfer
with bass instruments remaining the most challenging targets (RCA $>$ RPA, dominated by octave errors). These results
support the use of the proposed extractor as a robust front-end for rapid, instrument-specific conditioning in DDSP-style
neural synthesis and personalization.

\bibliographystyle{apalike}
{\small
\bibliography{ms.bib}}

\section*{\uppercase{Appendix}}

\subsection*{Dataset Details and Statistics}
\label{apd:dataset}

The research in this paper utilizes two distinct datasets: MedleyDB, which consists of real-world multitrack recordings, and MDB-stem-synth, a dataset of clean, resynthesized stems. As noted in the main paper, MedleyDB served as the source for acoustically complex training data (handling bleed, noise, and reverb), while MDB-stem-synth was used for evaluation against a clean, precisely annotated ground truth.

Table \ref{tab:dataset_stats} provides a detailed statistical comparison of the specific instrument subsets from both datasets used in this study. The comparison highlights the differences in the total number of clips, cumulative audio duration, and average clip length for each instrument.

\begin{table*}[h!]
\centering
\caption{Statistical comparison of instrument audio used from MedleyDB (real recordings) and MDB-stem-synth (synthesized stems). Durations are in seconds.}
\label{tab:dataset_stats}
\sisetup{
    table-format=5.2,  
    group-separator={,}, 
    detect-weight     
}
\begin{tabular}{
    l
    S[table-format=3.0]
    S[table-format=5.2]
    S[table-format=3.2]
    S[table-format=3.0]
    S[table-format=5.2]
    S[table-format=3.2]
}
\toprule
\multirow{2}{*}{\textbf{Instrument}} & \multicolumn{3}{c}{\textbf{MedleyDB (Recordings)}} & \multicolumn{3}{c}{\textbf{MDB-stem-synth (Synthesized)}} \\
\cmidrule(lr){2-4} \cmidrule(lr){5-7}
& {Clips} & {Total (s)} & {Avg. (s)} & {Clips} & {Total (s)} & {Avg. (s)} \\
\midrule

Clarinet & 9 & 885.39 & 98.38 & 8 & 829.21 & 103.65 \\
Cello & 13 & 2344.35 & 180.34 & 11 & 1816.80 & 165.16 \\
Double Bass & 17 & 2919.01 & 171.71 & 14 & 1722.81 & 123.06 \\
Electric Bass & 63 & 11398.80 & 180.93 & 58 & 9908.25 & 170.83 \\
Violin & 27 & 6986.28 & 258.75 & 14 & 3619.81 & 258.56 \\

\midrule
\textbf{Total} & {\textbf{129}} & {\textbf{24533.83}} & {--} & {\textbf{105}} & {\textbf{17896.88}} & {--} \\
\bottomrule
\end{tabular}
\end{table*}

For the selected instruments, the MedleyDB subset contains a larger clip count (129 vs. 105) and significantly more total audio data (approx. 24.5k seconds vs. 17.9k seconds). This provided a larger and more acoustically diverse corpus for self-supervised training. Notably, the average duration per clip is comparable for some instruments (e.g., violin, electric bass) while diverging on others (e.g., clarinet, double bass), reflecting the different compositions of the two datasets.

\subsection*{Training Objectives}
\label{apd:losses}

The total loss for a batch of $N$ frames is the average of the per-sample losses, where each component is defined below:
$$
\mathcal{L}_{\text{total}} = \frac{1}{N} \sum_{i=1}^{N} \left( w_i(\mathcal{L}_{\text{equiv},i} + \mathcal{L}_{\text{invar},i} + \mathcal{L}_{\text{sce},i}) \right)
$$
This formulation uses the soft weight $w_i$ to ensure that the $F_0$-specific equivariance and invariance losses are primarily driven by samples identified as strongly voiced.

\paragraph{Equivariance Loss for $F_0$ Estimation ($\mathcal{L}_{\text{equiv}}$).} To learn $F_0$, we enforce pitch equivariance. We define a set of weights $w_k = \alpha^k$ for $k \in \{p_{\text{min}}, \dots, p_{\text{max}}\}$, where $\alpha$ is a constant. If an input $X$ is pitch-shifted by $\delta$ bins to produce $X'$, the corresponding outputs $z$ and $z'$ should satisfy $z' \approx v^\delta z$. The loss is formulated using a symmetric Huber loss \cite{r-m:huber-1964}:
$$
\mathcal{L}_{\text{equiv}} = \frac{1}{2} \left( \mathcal{H}\left(\frac{z'}{z} - \alpha^\delta\right) + \mathcal{H}\left(\frac{z}{z'} - \frac{1}{\alpha^\delta}\right) \right)
$$

This loss enforces that transposing the input by $\delta$ bins produces a corresponding multiplicative shift in the intermediate representation $z$: specifically, $z' \approx \alpha^\delta z$, where $\alpha$ defines the exponential pitch scale matching the logarithmic CQT structure. The symmetric Huber loss provides robustness against transients and noisy frames while ensuring the model learns pitch intervals as consistent multiplicative factors without requiring labeled data.

\paragraph{Invariance Loss for Timbre Representation ($\mathcal{L}_{\text{invar}}$).} While $F_0$ is equivariant to pitch shifts, the instrumental timbre should be invariant. Given an original output $y$ and the output from a pitch-shifted version $y'$, the timbre representation should be similar. This is achieved using a symmetric cross-entropy loss \cite{r:beigi-sr-book-2011}:
$$
\mathcal{L}_{\text{invar}} = \frac{1}{2} \left( \text{CE}(y, p') + \text{CE}(y', p) \right)
$$

This loss encourages the model to learn pitch-invariant timbre features by enforcing similarity between output logits $y$ and $y'$ from original and pitch-shifted inputs. Since timbre characteristics such as spectral shape, harmonic decay rates, and formants remain consistent across pitches, the symmetric cross-entropy loss penalizes differences in the predicted distributions. This disentanglement of pitch from timbre is essential for downstream tasks including instrument identification, synthesis control, and timbre transfer, ensuring learned features generalize across different notes from the same instrument.

\paragraph{Shift Cross-Entropy Loss ($\mathcal{L}_{\text{sce}}$).} This loss enforces a circular equivariance property and is the key signal for the E-step. If an input is pitch-shifted by $\delta$ bins, the output distribution should be circularly shifted by the same amount. Let $y$ be the original output and $y'$ be the transformed output. The loss is \cite{Riou2023PESTO}:
$$
\mathcal{L}_{\text{sce}} = \frac{1}{2} \left( \text{CE}(y', \text{roll}(p, \delta)) + \text{CE}(y, \text{roll}(p', -\delta)) \right)
$$

This core self-supervised loss enforces circular equivariance: if the input is transposed by $\delta$ bins, the output probability distribution should be circularly shifted by the same amount. It compares the predicted distribution from the shifted input with the appropriately rolled version of the original distribution using symmetric cross-entropy. This constraint refines $F_0$ prediction by ensuring the entire predicted pitch distribution shifts correctly and consistently, encouraging sharp, well-localized probabilities centered on the correct bin. Critically, $\mathcal{L}_{\text{sce}}$ also serves as an implicit voicing detector: frames with clear harmonic structure produce sharp distributions that shift predictably, yielding low SCE loss, while noisy or unvoiced frames generate flatter, unstable distributions with high SCE loss. This property enables the E-step of the iterative re-weighting algorithm to automatically identify and down-weight unreliable frames, making $F_0$ learning robust to real-world noise and artifacts as detailed in .

\subsection*{Voicing Classifier}
\label{apd:voicing}

After the self-supervised representation learning phase is complete, the weights of the network $f(\cdot)$ are frozen. The final set of sample weights, $w_i$, computed for the entire training dataset during the iterative re-weighting, are used to generate hard pseudo-labels. These weights, which range from 0 to 1, effectively capture the model's converged confidence that a frame contains a stable, voiced signal. A hard pseudo-label $v_i^* \in \{0, 1\}$ is assigned by thresholding these final weights:
$$
v_i^* = \begin{cases} 1 & \text{if } w_i > \theta \\ 0 & \text{otherwise} \end{cases}
$$
where $\theta$ is a predefined confidence threshold. These pseudo-labels serve as the targets for the subsequent supervised training.

\begin{figure*}[!ht]
    \centering
    \includegraphics[width=1\linewidth]{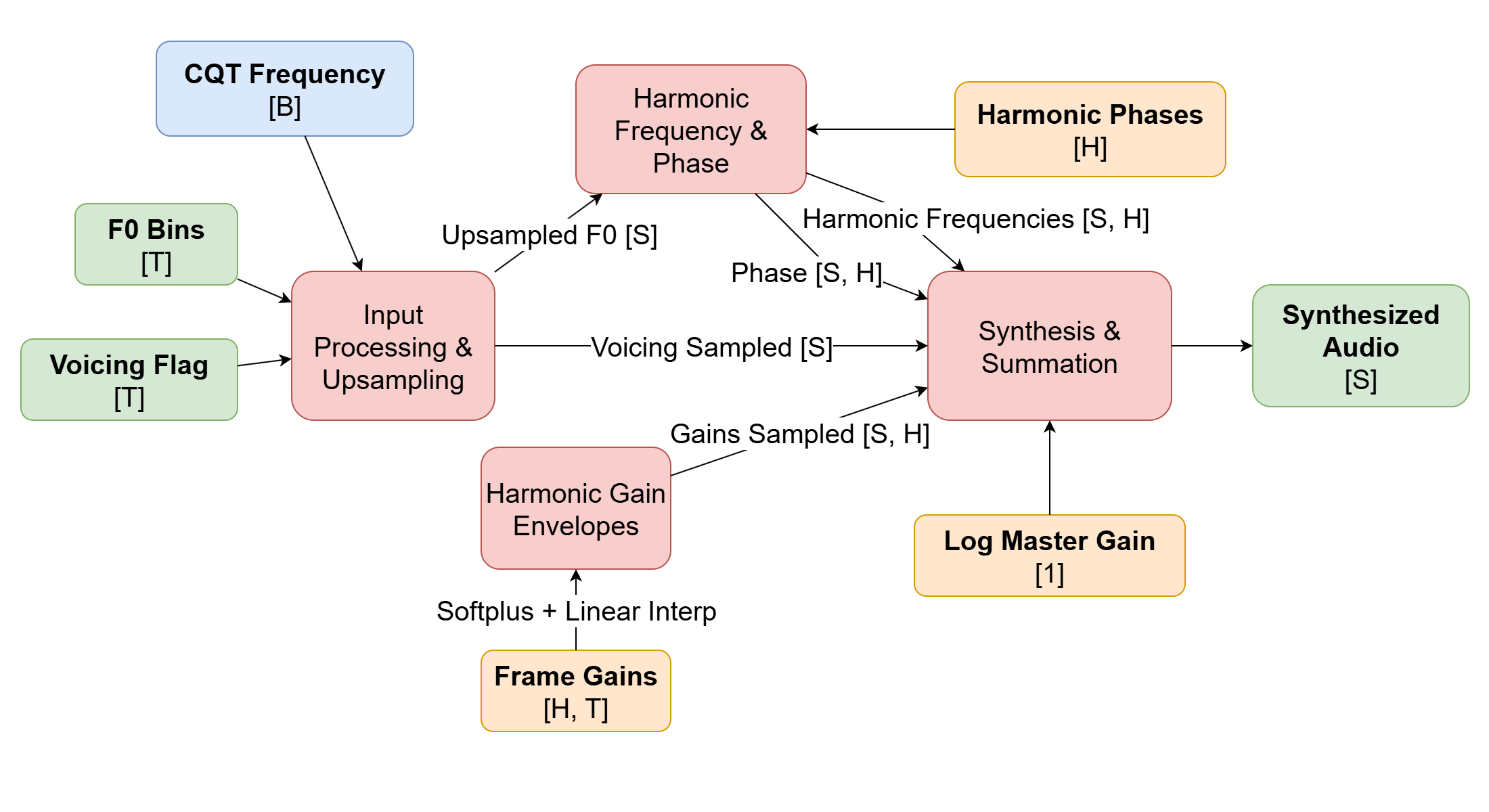}
    \caption{Harmonic Synthesizer Architecture for Analysis-by-Synthesis. Inputs (`$F_0$ Bins', `Voicing Flag') are processed and upsampled. Learnable parameters (`Frame Gains', `Harmonic Phases', `Log Master Gain') control harmonic frequencies, phases, and amplitudes. The synthesis stage combines these components, applying voicing and Nyquist masking, to generate the final audio output.}
    \label{fig:synth_diagram}
\end{figure*}

A lightweight, single-layer linear classification head is trained directly on the CQT spectrograms. This layer maps each CQT input frame $X$ to a single logit for binary (voiced/unvoiced) classification. The classifier is optimized using standard binary cross-entropy loss against the generated pseudo-labels $v^*$. We deliberately choose a small, single-layer architecture over deeper multi-layer networks because the EM re-weighting process inherently down-weights some genuinely voiced frames during early iterations when $F_0$ estimates are still inaccurate. Larger models with greater capacity tend to overfit to these imperfect pseudo-labels, incorrectly learning to classify initially down-weighted but actually voiced frames as unvoiced. The limited capacity of a single linear layer provides natural regularization, forcing the classifier to capture only the most robust voicing patterns while generalizing better to frames that were temporarily misclassified during the iterative training process. While the feature extractor $f(\cdot)$ is not used for inference in this stage, its role in generating the high-quality pseudo-labels through the EM procedure is critical. This two-stage approach leverages the pitch structure learned during self-supervision to train a final, high-performance voicing detector without requiring any external ground-truth labels.

\subsection*{Harmonic Synthesizer}
\label{apd:harmonic}

To qualitatively validate the accuracy of our predicted fundamental frequency ($\hat{F}_{0}$) and voicing ($v$) contours, we employ an analysis-by-synthesis procedure. We use a differentiable harmonic synthesizer, illustrated in Figure \ref{fig:synth_diagram}, to reconstruct an audio segment using only the predicted $\hat{F}_{0}$ and $v$ as inputs. The synthesizer's parameters are then optimized to match the target audio, serving as a robust test of whether the predicted contours are sufficient for high-fidelity reconstruction.

\paragraph{Model Architecture}
The synthesizer is a specialized additive model that generates audio by summing $H$ harmonics, where the amplitude of each harmonic is a learnable parameter that varies over time. The model's inputs are the frame-wise $\hat{F}_{0}$ bin indices and the binary voicing flags $v$, both at the CQT frame rate (hop length of 160 samples).

\begin{enumerate}
    \item \textbf{Input Pre-processing:} The $\hat{F}_{0}$ bin indices are first converted to Hertz using the CQT frequency lookup table. Both the $\hat{F}_{0}$ (Hz) contour and the voicing contour $v$ are upsampled from the frame rate to the audio sample rate ($f_s = 16000\text{ Hz}$). We use linear interpolation for the $\hat{F}_{0}$ contour to ensure smooth frequency transitions and nearest-neighbor interpolation for the voicing contour to maintain sharp onsets and offsets. This results in sample-rate signals $f_0(t)$ and $v(t)$.

    \item \textbf{Instantaneous Phase:} The instantaneous frequency for the $h$-th harmonic is $f_h(t) = h \cdot f_0(t)$. The instantaneous phase $\phi_h(t)$ is calculated by integrating the frequency, approximated as a cumulative sum of the phase increment, and adding a learnable per-harmonic global phase offset $\psi_h$:
    $$
    \phi_h(t) = \left( 2\pi \sum_{i=1}^{t} \frac{h \cdot f_0(i)}{f_s} \right) + \psi_h
    $$
    where $f_s$ is the sample rate (16000 Hz) and $\psi \in \mathbb{R}^{H}$ is a learnable parameter vector.

    \item \textbf{Learnable Amplitude Envelopes:} The core of the synthesizer is its time-varying amplitude control. A learnable parameter matrix $G_{\text{pre}} \in \mathbb{R}^{H \times T}$ stores a pre-activation gain for each of the $H$ harmonics at each of the $T$ frames. These gains are passed through a \texttt{Softplus} function to ensure positivity, $G_{\text{f}} = \text{Softplus}(G_{\text{pre}})$, and are then linearly interpolated to the sample rate, yielding $g_h(t)$.

    \item \textbf{Synthesis:} The final audio signal $s(t)$ is generated by summing the harmonics, each modulated by its specific gain envelope, the master voicing gate $v(t)$, and a learnable logarithmic master gain $g_{\text{master}}$. Harmonics with frequencies above the Nyquist limit ($f_s / 2$) are masked out.
    $$
    s(t) = 10^{g_{\text{master}}} \cdot v(t) \cdot \sum_{h=1}^{H} g_h(t) \cdot \sin(\phi_h(t))
    $$
\end{enumerate}

\paragraph{Optimization}
For a given target audio segment $y(t)$ and its corresponding $\hat{F}_{0}$ and $v$ contours, we optimize the model's learnable parameters: the frame-wise gains $G_{\text{pre}}$, the harmonic phases $\psi$, and the master gain $g_{\text{master}}$.

The objective is to minimize the perceptual difference between the synthesized audio $s(t)$ and the target $y(t)$. We use the Multi-Resolution STFT Loss (MR-STFT) from the \texttt{auraloss} library, which computes the L1 loss on the magnitudes and L2 loss on the log-magnitudes of the STFT across multiple resolutions (FFT sizes, hop sizes, and window sizes).

To encourage smooth and realistic amplitude envelopes, we add a smoothness regularizer. This penalty is the mean temporal Total Variation (TV) of the frame-wise gain matrix $G_{\text{f}}$, penalizing large, abrupt changes between adjacent frames:
$$
\mathcal{L}_{\text{smooth}} = \frac{1}{H(T-1)} \sum_{h=1}^{H} \sum_{t=1}^{T-1} | G_{\text{f}}[h, t+1] - G_{\text{f}}[h, t] |
$$
The final loss is a weighted sum:
$$
\mathcal{L}_{\text{total}} = \mathcal{L}_{\text{MR-STFT}}(s, y) + \lambda \cdot \mathcal{L}_{\text{smooth}}
$$
The model is trained for a fixed number of epochs using the Adam optimizer. This process effectively performs analysis-by-synthesis, finding the optimal harmonic amplitudes required to reconstruct the target audio given our model's $\hat{F}_{0}$ and voicing predictions.

\end{document}